\documentclass[11pt]{article}

\newsavebox{\foobox}
\newcommand{\slantbox}[2][0]{\mbox{%
        \sbox{\foobox}{#2}%
        \hskip\wd\foobox
        \pdfsave
        \pdfsetmatrix{1 0 #1 1}%
        \llap{\usebox{\foobox}}%
        \pdfrestore
}}
\newcommand\unslant[2][-.25]{\slantbox[#1]{$#2$}}

\newcommand{\mpi}{\text{\unslant[-.18]\pi}}
\newcommand{\mdelta}{\text{\unslant[-.18]\delta}}

\usepackage[left=2cm, right=2cm, top=2.5cm, bottom=2.5cm]{geometry}
\usepackage[x11names]{xcolor}
\usepackage{fancyhdr, amssymb, cancel, amsmath, graphicx, pgfplots, tikz}
\usepackage{isomath}

\usetikzlibrary{shadows}

\newcommand{\stylecolor}{blue!50!black}

\usepackage[labelfont={bf,sf, color=\stylecolor}, margin={1.5cm,0cm}]{caption}

\usepackage[colorlinks=true, urlcolor=\stylecolor!70!white, linkcolor=\stylecolor, citecolor=\stylecolor!70!white, hyperindex=true, linktocpage=true]{hyperref}

\usepackage[explicit]{titlesec}

\newcommand*\sectionlabel{}
\titleformat{\section}
  {\gdef\sectionlabel{}
   \Large\bfseries\scshape}
  {\gdef\sectionlabel{\thesection }}{0pt}
  {\begin{tikzpicture}[remember picture,overlay]
       \end{tikzpicture}
  }
\titlespacing*{\section}{0pt}{0pt}{0pt}

\newcommand*\subsectionlabel{}
\titleformat{\subsection}
  {\gdef\subsectionlabel{}
   \large\bfseries\scshape}
  {\gdef\subsectionlabel{\thesubsection  }}{0pt}
  {\begin{tikzpicture}[remember picture]
    	\draw (-0.15, 0) node[left] {\color{\stylecolor} \textsf{\subsectionlabel}};
	\draw (0.15, 0) node[right] {\color{\stylecolor} \textsf{#1}};
	\fill[color=\stylecolor] (-0.05, -0.23) rectangle (0.05, 0.23);
       \end{tikzpicture}
  }
\titlespacing*{\subsection}{-4pt}{10pt}{0pt}

\newcommand*\subsubsectionlabel{}
\titleformat{\subsubsection}
  {\gdef\subsubsectionlabel{}
   \bfseries\scshape}
  {\gdef\subsubsectionlabel{\thesubsubsection.\ \  }}{0pt}
  {\begin{tikzpicture}[remember picture]
    	\draw (0, 0) node[left] {\color{\stylecolor} \textsf{\subsubsectionlabel}};
	\draw (0, 0) node[right] {\color{\stylecolor} \textsf{#1}};
       \end{tikzpicture}
  }
\titlespacing*{\subsubsection}{-4pt}{7pt}{0pt}

\pgfplotsset{every axis legend/.append style={at={(1.02,1)},anchor=north west}}

\graphicspath{{figures/}}

\begin{document}

\allowdisplaybreaks

\pagestyle{fancy}
\renewcommand{\headrulewidth}{0pt}
\fancyhead{}

\fancyfoot{}
\fancyfoot[C] {\textsf{\textbf{\thepage}}}

\begin{equation*}
\begin{tikzpicture}
\draw (\textwidth, 0) node[text width = \textwidth, right] {\color{white} easter egg};
\end{tikzpicture}
\end{equation*}

\begin{equation*}
\begin{tikzpicture}
\draw (0.5\textwidth, -3) node[text width = \textwidth] {\huge  \textsf{\textbf{Electronic sound modes and plasmons in  \\ \vspace{0.07in}hydrodynamic two-dimensional metals  }} };
\end{tikzpicture}
\end{equation*}
\begin{equation*}
\begin{tikzpicture}
\draw (0.5\textwidth, 0.1) node[text width=\textwidth] {\large \color{black}  \textsf{Andrew Lucas}$^{\color{\stylecolor} \mathsf{a}}$ \textsf{and  Sankar Das Sarma}$^{\color{\stylecolor} \mathsf{b}}$ };
\draw (0.5\textwidth, -0.5) node[text width=\textwidth] {$^{\color{\stylecolor} \mathsf{a}}$ \small{\textsf{Department of Physics, Stanford University, Stanford, CA 94305, USA}}};
\draw (0.5\textwidth, -0.6) node[text width=\textwidth, below] {$^{\color{\stylecolor} \mathsf{b}}$ {\small \textsf{Condensed Matter Theory Center and Joint Quantum Institute, Department of Physics, University of Maryland, College Park, MD 20742 USA}}};
%\draw (0.5\textwidth, -0.5) node[text width=\textwidth] {$^{\color{\stylecolor} \mathsf{a}}$ \small\textsf{present address: Technische Universit\"at Dresden, Dresden, Germany}};
\end{tikzpicture}
\end{equation*}
\begin{equation*}
\begin{tikzpicture}
\draw (0, -13.1) node[right, text width=0.5\paperwidth] {\texttt{ajlucas@stanford.edu}};
\draw (\textwidth, -13.1) node[left] {\textsf{\today}};
\end{tikzpicture}
\end{equation*}
\begin{equation*}
\begin{tikzpicture}
\draw[very thick, color=\stylecolor] (0.0\textwidth, -5.75) -- (0.99\textwidth, -5.75);
\draw (0.12\textwidth, -6.25) node[left] {\color{\stylecolor}  \textsf{\textbf{Abstract:}}};
\draw (0.53\textwidth, -6) node[below, text width=0.8\textwidth, text justified] {\small Using an analytically tractable kinetic model of a two dimensional Fermi liquid of electrons, we characterize the crossovers between zero sound, first sound and plasmons.  For experimentally realized Fermi liquids in a hydrodynamic limit, both zero and first sound waves are essentially replaced by plasmons.    The plasmon dispersion relation is robust against hydrodynamic effects, up to acquiring the viscous-limited decay rate of a first sound wave in the hydrodynamic limit.  We discuss implications for experiments in clean two dimensional electron gases.};
\end{tikzpicture}
\end{equation*}

\tableofcontents

\begin{equation*}
\begin{tikzpicture}
\draw[very thick, color=\stylecolor] (0.0\textwidth, -5.75) -- (0.99\textwidth, -5.75);
\end{tikzpicture}
\end{equation*}

\titleformat{\section}
  {\gdef\sectionlabel{}
   \Large\bfseries\scshape}
  {\gdef\sectionlabel{\thesection }}{0pt}
  {\begin{tikzpicture}[remember picture]
	\draw (0.2, 0) node[right] {\color{\stylecolor} \textsf{#1}};
	\draw (0.0, 0) node[left, fill=\stylecolor,minimum height=0.27in, minimum width=0.27in] {\color{white} \textsf{\sectionlabel}};
       \end{tikzpicture}
  }
\titlespacing*{\section}{0pt}{20pt}{5pt}

\section{Introduction}
Advances in  the quality of crystal growth have led to strong experimental evidence for the hydrodynamic flows of electrons in solid state systems  \cite{bandurin, crossno, mackenzie, levitov1703};  see \cite{lucasreview17} for  a recent review.   Recent theoretical work  has clarified the  signatures of hydrodynamic behavior on correlated electron flow in a diverse  set of materials \cite{andreev, lucas3, polini, levitovhydro, levitov1607, lucas1612, levitov1612, hartnoll1705, lucasRFB}, including those where the textbook Navier-Stokes equations are not applicable.

The condition for an electron liquid to obey hydrodynamics was discussed in the literature a long time ago \cite{pines, landauvol9, giulianivignale}.  The key condition is that electron-electron (or equivalently, inter-particle) collision rate must be rapid enough to bring about local thermal equilibrium.  This necessary condition boils down (for an electron liquid, the subject of interest in the current paper) to the electron-electron interaction induced scattering rate being larger than the momentum relaxation scattering rates associated with electron-impurity and/or electron-phonon scattering rates.  Leaving aside phonon scattering, which is usually important only at rather high temperatures, this implies that electron-electron scattering should be stronger than electron-impurity scattering.  Since electron-electron scattering rates are typically $\propto T^2$ in a Fermi liquid, it may seem that at some finite temperature, all metals should manifest hydrodynamic behavior.  This turns out to be untrue, and typically in normal 3D metals (even in the cleanest possible scenario) electron-electron scattering rate is rather weak, and the hydrodynamic condition is never satisfied.  On the other hand, graphene and ultra-clean 2D GaAs-based systems do satisfy the hydrodynamic condition at moderate temperatures.  

%One of the most basic predictions of a theory of hydrodynamics is the emergence of a collective sound wave.   Such a first sound mode has been experimentally observed in the hydrodynamic regime of the Fermi liquid He-3 \cite{abel}.  To date, there has been no experimental  observation of a sound wave of electrons in a solid state system, even  those which are expected to be  in the hydrodynamic regime.     

An immediate consequence of hydrodynamics is the existence of an electronic sound mode (the so-called ``first sound") where the long wavelength energy dispersion of the collective mode is linear in wave number:  $\omega \propto k$.  This collective electronic sound mode has never been observed in any 2D electron liquid (either graphene or 2D semiconductor systems).   Even an indirect observation of sound waves through the  hydrodynamic Dyakonov-Shur instability  \cite{DS} has never been cleanly observed, despite many years of experimental efforts \cite{tauk, giliberti}.     Instead, the observed collective mode is the usual long wavelength plasmon mode, with $\omega \propto \sqrt{k}$ due to the long-range nature of the Coulomb interaction.  As we will see, this may be understood as an appropriate 'zero sound' mode in the collisionless regime.  This casts a shadow on the theoretically proposed hydrodynamic descriptions of strongly interacting electron liquid, as a fluid should have a sound mode.  Our work aims to clearly resolve this conundrum.   Appropriately incorporating long-range Coulomb interactions into a hydrodynamic and kinetic model, we show that hydrodynamics in 2D metals is consistent with a long wavelength collective mode dispersing as $\omega \propto \sqrt{k}$.   For purely short-range interactions, there is indeed a linear-in-wavenumber first sound mode in the hydrodynamic regime (as observed, for example, in normal He-3 \cite{abel}), but Coulomb interaction modifies this dispersion to a square-root in wavenumber dependence at long wavelength.

\subsection{Summary of Results}
In this paper, we present a simple and analytically tractable kinetic theory model for the dynamics of a two dimensional Fermi liquid.  We account for both Landau's Fermi liquid interaction function, and the long-range nature of the Coulomb interaction, as appropriate for two-dimensional metals. We observe that in a typical Fermi liquid with strong electron-electron interactions, both  the first and zero sound mode are replaced by a plasmon  mode which persists (at least) until the breakdown of the kinetic theory.    The main effects of hydrodynamics on this conventional plasmon mode are in the decay rate  of the excitation, which is  more challenging to experimentally measure.  Furthermore, the higher-order wave-number corrections to the collective mode dispersions are also affected by hydrodynamics, and can be captured in our model (the leading-order dispersion is fixed by the Coulomb interaction to be a plasmonic $\sqrt{k}$).

Let us be more quantitative.   We study a toy model of an isotropic two dimensional Fermi liquid with a  single \color{black} electron-electron scattering rate $\gamma $,  electron-impurity (momentum-relaxing) scattering rate $\gamma_{\mathrm{imp}} \le \gamma  $,  and effective interaction constant $\alpha$.     We consider spinless fermions, in order to focus our discussion the salient features of plasmon physics. The generalization to the spinful case is straightforward.    \color{black}  In a Fermi liquid $\gamma \sim T^2$  (up to logarithms \cite{sarma96,  novikov, sarma13, polini1506}).  We also take the only non-vanishing Landau parameter of Fermi liquid theory to be $\mathcal{F}_0$, for the sake of simplicity (although a generalization including more Landau parameters is straightforward, but cumbersome).     As we have assumed spinless fermions, $\mathcal{F}_0$ corresponds to a Landau parameter in the symmetric channel.  \color{black} First, let us assume the absence of long-range Coulomb interactions (which can be experimentally achieved by placing gates very close to the sample to screen out  the long-range part of the interaction).  If $\gamma_{\mathrm{imp}}=0$, then we find a simple crossover between hydrodynamic first sound waves, with  dispersion relation 
\begin{equation}
    \omega = \pm \sqrt{\frac{1+\mathcal{F}_0}{2}} v k  - \mathrm{i}\frac{v^2}{8\gamma}k^2 + \mathrm{O}(k^3),  \label{eq:puresound}
  \end{equation}
  valid when $|\omega| \ll \gamma$, and collisionless zero sound when $|\omega| \gg \gamma$: \begin{equation}
 \omega =  \pm \frac{1+\mathcal{F}_0}{\sqrt{1+2\mathcal{F}_0}} v k -\mathrm{i}\gamma \frac{1+\mathcal{F}_0}{(1+2\mathcal{F}_0)^2} +  \mathrm{O}\left(\frac{1}{k}\right).  \label{eq:zerosound}
  \end{equation}
Whenever $\mathcal{F}_0>0$,  the speed of zero sound $v_{\mathrm{s0}}$ is always faster than the speed of first sound $v_{\mathrm{s1}}$:  \begin{equation}
 1 \le  \frac{v_{\mathrm{s0}}}{v_{\mathrm{s1}}}  = \sqrt{\frac{2+2\mathcal{F}_0}{1+2\mathcal{F}_0}}  \le  \sqrt{2}.
  \end{equation}
  So the most important difference between these two sound waves (beyond their physical interpretation) is the decay rate:  $\mathrm{Im}(\omega) \sim T^{-2}$ in the hydrodynamic limit, while $\mathrm{Im}(\omega) \sim T^2$ in the collisionless limit.  These results are qualitatively similar to the well-understood theory of first and zero sound waves in  He-3, which were experimentally  observed a long time ago \cite{abel}.  

If $\gamma_{\mathrm{imp}} \ne 0$, then sound waves and plasmons are destroyed at long wavelengths:  impurity collisions act as  a cutoff, overdamping these ballistically propagating modes at low enough frequency. The  two sound modes merge as $k\rightarrow 0$, and form a purely diffusive  mode associated with the Ohmic diffusion of charge:   \begin{equation}
\omega = -\mathrm{i}\frac{v^2_{\mathrm{s1}}}{\gamma_{\mathrm{imp}}}k^2  \label{eq:omegaimp}
\end{equation} as well as a gapped mode associated with momentum relaxation:  $\omega \approx -\mathrm{i}\gamma_{\mathrm{imp}}$.    Eq. (\ref{eq:omegaimp}) defines the well-known diffusion pole of the disordered electron Green's function.  When $v_{\mathrm{s1}}k \gtrsim \gamma_{\mathrm{imp}}$, we instead find the dispersion relation of sound waves described above, but with the decay rates approximately shifted by \begin{equation}
\mathrm{Im}(\omega;  \gamma_{\mathrm{imp}}\ne 0) \approx \mathrm{Im}(\omega;  \gamma_{\mathrm{imp}}= 0) - \frac{\gamma_{\mathrm{imp}}}{2}.  \label{eq:soundoffset}
\end{equation}
For zero sound waves, the shift is slightly more  complicated -- see Eq. (\ref{eq:zerosoundimp}) below.

Our most important result is that when we include long-range Coulomb interactions, we can approximately replace \begin{equation}
\mathcal{F}_0 \rightarrow \mathcal{F}_0 + \frac{2\mpi\alpha}{\lambda_{\mathrm{F}}|k|},  \label{eq:F0plasmon}
\end{equation}
with $\lambda_{\mathrm{F}}$ the Fermi wavelength.    This may be construed as an effective generalization to a Landau-Silin theory applicable to 2D metals, as compared with a 2D Fermi liquid theory appropriate for neutral Fermi liquids.     Eq. (\ref{eq:F0plasmon}) can also be understood from the RPA approximation \cite{zala2001}. The $1/k$ dependence in Eq. (\ref{eq:F0plasmon}) is the precise 2D Coulomb interaction behavior for the $1/r$ long-range potential.    We then find that first sound (\ref{eq:puresound}) is replaced by \begin{equation}
\omega \approx \pm \sqrt{\frac{\mpi \alpha v^2}{\lambda_{\mathrm{F}}}|k|} - \frac{\mathrm{i}v^2}{8\gamma}k^2 + \mathrm{O}\left(k^3\right),  \label{eq:plasmon1}
\end{equation}
and the collisionless zero sound is  replaced by  \begin{equation}
\omega \approx \pm \sqrt{\frac{\mpi \alpha v^2}{\lambda_{\mathrm{F}}}|k|} -\mathrm{i}  \gamma  \frac{\lambda_{\mathrm{F}} |k|}{8\mpi\alpha} .  \label{eq:plasmon2}
\end{equation}
   The quantitative prefactor of $\mathrm{Re}(\omega)$ in \emph{both} of the above equations is identical to the well known plasmon dispersion relation \cite{stern, hwangplasmon, sarmaplasmon}.  The main results of this paper are twofold.  First, we derive the unexpected $\mathrm{Im}(\omega) \sim |k|$ scaling for the decay rate of plasmons in a collisionless regime.   Second, we show that the crossover between the ``hydrodynamic plasmon"  (\ref{eq:plasmon1}) and the ``collisionless plasmon" (\ref{eq:plasmon2}) is when \begin{equation}
k \sim \frac{\lambda_{\mathrm{F}} \gamma^2}{\alpha  v^2}.  \label{eq:plasmontrans}
\end{equation}   
This crossover occurs at a much smaller wave number than naively expected ($k\sim \gamma/v$).     The  breakdown of these equations,  and the replacement of plasmons with conventional first or zero sound modes, occurs when \begin{equation}
\lambda_{\mathrm{F}}k \gtrsim \frac{2\mpi\alpha}{\max(1,\mathcal{F}_0)}
\end{equation}
which is generally outside the regime of validity of kinetic theory.  \color{black} Since $\alpha \gtrsim  1$  is indeed required to observe hydrodynamic behavior in a two-dimensional Fermi liquid  at present, it is not possible to observe either a first or zero  sound mode:   the long-range Coulomb  interactions will destroy conventional sound waves at all wavelengths, and the only observable long wavelength collective mode would be the usual plasmon mode, which indeed seems to be the generic experimental situation.  Thus, in contrast to the 3D neutral Fermi liquid He-3, the hydrodynamic behavior in 2D Coulomb Fermi systems may not necessarily manifest long wavelength sound modes.

These results clarify previous literature relating plasmons to sound waves in Fermi liquids.   In particular, the hydrodynamic and collisionless plasmon are essentially the same mode at  long wavelengths in a Fermi liquid, albeit with distinct decay mechanisms (and different subleading wave-number corrections).  A crossover between plasmons and first sound will only be observable in nearly charge neutral (non-degenerate) systems, such as the Dirac fluid in graphene \cite{sarma13, levitovsound, fogler3,  lucasplasma}.

\section{The Boltzmann Equation}
In order to derive these results, we first review the kinetic theory of a two dimensional Fermi liquid with a circular Fermi surface.   
Within linear response, the most important changes to the distribution function $f$ occur at the Fermi surface:
 \begin{equation}
f \approx f_{\mathrm{eq}} + \mdelta f \equiv \mathrm{\Theta}(\mu - \epsilon(\mathbf{p})) + \mdelta(\mu - \epsilon(\mathbf{p})) \Phi(\mathbf{x},\theta,t).
\end{equation}  
Here $\tan \theta = p_y/p_x$ denotes the angle on the circular Fermi surface.   In what follows, we will denote with $p_{\mathrm{F}}$ the Fermi momentum $\epsilon(p_{\mathrm{F}}) = \mu$, and also define the Fermi velocity \begin{equation}
v \equiv  \epsilon^\prime(p_{\mathrm{F}}).
\end{equation}
The quantum Boltzmann equation,  neglecting  spin, \color{black} reads
\begin{equation}
\frac{\partial f}{\partial t}  + \frac{\partial \epsilon}{\partial \mathbf{p}} \cdot \frac{\partial f}{\partial \mathbf{x}} - \frac{\partial \epsilon}{\partial \mathbf{x}} \cdot \frac{\partial f}{\partial \mathbf{p}}   = \mathcal{C}[f]  \label{eq:Fboltz}
\end{equation}
 $\mathcal{C}[\Phi]$ is the collision integral, and $\epsilon$ denotes the energy of a quasiparticle, accounting for both the bare band structure $\epsilon_0(\mathbf{p})$, Landau's short-range interaction function $\mathcal{E}$, and the long-range part of the Coulomb  interaction, arising from fluctuations in the Fermi surface:  \begin{equation}
 \epsilon(\mathbf{x},\mathbf{p}) = \epsilon_0(\mathbf{p}) + \mathcal{E}[f(\mathbf{x},\mathbf{p})] + \frac{e^2}{4\mpi \varepsilon} \int  \frac{\mathrm{d}^2\mathbf{y} \mathrm{d}^2\mathbf{q}}{(2\mpi\hbar)^2} \;  \frac{f(\mathbf{x},\mathbf{p})(f(\mathbf{y},\mathbf{q}) - f_{\mathrm{eq}})}{|\mathbf{x}-\mathbf{y}|}.  \label{eq:epsilon}
 \end{equation}
We have  treated Coulomb interactions here in the conventional self-consistent Vlasov approximation.    The interaction function is \begin{equation}
\mathcal{E}[\Phi] \approx \int\limits_{-\mpi}^{\mpi}  \frac{\mathrm{d}\phi}{2\mpi} \mathcal{F}(\theta-\phi)\Phi(\phi),
\end{equation} 
with $\mathcal{F}$ an  even function.   Combining Eqs. (\ref{eq:Fboltz}) and (\ref{eq:epsilon}),  we obtain \begin{equation}
\frac{ \partial  \mdelta f}{\partial t} + \frac{\partial \epsilon_0}{\partial \mathbf{p}} \cdot \frac{\partial \mdelta f}{\partial \mathbf{x}} - \frac{\partial f_{\mathrm{eq}}}{\partial \mathbf{p}} \cdot \frac{\partial \epsilon}{\partial \mathbf{x}} = \mathcal{C}[f],
 \end{equation}
 which leads to 
 \begin{equation}
\frac{\partial \Phi}{\partial t}  + \mathbf{v}\cdot\left[ \frac{\partial \Phi}{\partial \mathbf{x}}  + \frac{\partial \epsilon}{\partial \mathbf{x}}\right]   = \mdelta \mathcal{C}[\Phi]  \label{eq:PhiBoltz}
\end{equation}
where $\mathbf{v} = \partial \epsilon_0/\partial \mathbf{p}$ and  $\mdelta \mathcal{C}[\Phi]$ denotes a linearized collision integral, which will be a local in $\mathbf{x}$, but nonlocal in $\theta$, linear expression in $\Phi$.

In this paper we are interested in studying propagating waves.   So without loss of generality we may look for solutions to (\ref{eq:PhiBoltz}) of the form \begin{equation}
\Phi(\mathbf{x}, \theta,t) = \sum_{n\in\mathbb{Z}} a_n \mathrm{e}^{\mathrm{i}(\mathbf{k}\cdot\mathbf{x}-\omega t + n\theta)}.
\end{equation}
Writing \begin{equation}
\mathcal{F}(\phi) = \sum_{k\in\mathbb{Z}} \mathcal{F}_n \mathrm{e}^{\mathrm{i}n\phi},
\end{equation}
with $\mathcal{F}_n = \mathcal{F}_{-n}$, we find 
\begin{equation}
\frac{\partial \mathcal{E}[\Phi]}{\partial  \mathbf{x}} = \mathrm{i} \mathbf{k} \sum_{m,n} \int\limits_0^{2\mpi} \frac{\mathrm{d}\phi}{2\mpi} \mathcal{F}_m \mathrm{e}^{\mathrm{i}m(\theta-\phi)} a_n \mathrm{e}^{\mathrm{i}(\mathbf{k}\cdot \mathbf{x}-\omega t  +n\phi)} = \mathrm{i} \mathbf{k} \sum_n \mathcal{F}_n a_n \mathrm{e}^{\mathrm{i}(\mathbf{k}\cdot \mathbf{x}-\omega t  +n\phi)} \label{eq:Fn}
\end{equation}
Taking the spatial Fourier transform of the long-range Coulomb interaction, we find 
\begin{equation}
\frac{e^2}{4\mpi \varepsilon} \int  \mathrm{d}^2\mathbf{x}  \frac{\partial}{\partial \mathbf{x}} \int  \frac{\mathrm{d}^2\mathbf{y} \mathrm{d}^2\mathbf{q}}{(2\mpi\hbar)^2} \;  \frac{\mdelta(\mu - \epsilon(\mathbf{q}))\Phi(\mathbf{y},\theta_{\mathbf{q}},t)}{|\mathbf{x}-\mathbf{y}|} = \mathrm{i}\mathbf{k}  \frac{\alpha}{\lambda_{\mathrm{F}}|\mathbf{k}|}    a_0 \mathrm{e}^{\mathrm{i}(\mathbf{k}\cdot \mathbf{x}-\omega t)} \label{eq:a0}
\end{equation}
where \begin{equation}
\alpha \equiv \frac{e^2}{4\mpi \varepsilon \hbar v_{\mathrm{F}}} 
\end{equation}
is the effective fine structure constant, and $\lambda_{\mathrm{F}} = 2\mpi\hbar/p_{\mathrm{F}}$ is the Fermi wavelength.   Finally, similarly to (\ref{eq:Fn}),  rotational  invariance demands that \begin{equation}
\mdelta \mathcal{C}[\Phi]  = -\sum_n \gamma_n a_n \mathrm{e}^{\mathrm{i}(\mathbf{k}\cdot \mathbf{x}-\omega t  +n\phi)}  \label{eq:Cn}
\end{equation}
with $\gamma_n \ge 0$ also required by the second law of thermodynamics.   
Combining (\ref{eq:PhiBoltz}), (\ref{eq:Fn}), (\ref{eq:a0}) and (\ref{eq:Cn}), and choosing $\mathbf{k} = k\hat{\mathbf{x}}$, we obtain   \begin{equation}
-\mathrm{i}\omega a_n +  \frac{\mathrm{i}kv}{2}(a_{n+1}+a_{n-1}) + \frac{\mathrm{i}kv}{2} \left(\mathcal{F}_{n+1}a_{n+1} + \mathcal{F}_{n-1}a_{n-1}\right)  +  \mathrm{i} k \frac{\mpi v\alpha}{\lambda_{\mathrm{F}}|k|} a_0  \mdelta_{|n|,1} = - \gamma_n a_n. \label{eq:anfourier}
\end{equation}
This infinite set of algebraic equations governs the normal modes of our kinetic theory.

A key observation is that the replacement (\ref{eq:F0plasmon}) ``removes" the third term of (\ref{eq:anfourier}).  In other words, long-range Coulomb interactions are a $k$-dependent $\mathcal{F}_0$.   Since $k$ is  not dynamical in (\ref{eq:anfourier}),  we can therefore solve (\ref{eq:anfourier}) without explicitly accounting for long-range Coulomb interactions, and then include them at the end of the calculation through (\ref{eq:F0plasmon}).  Indeed, in the section that follows, we will set $\alpha=0$ and characterize the zero-to-first sound crossover of this model, which is akin to solving for the  collective modes in the short-range interacting 2d Fermi liquids.

Let us also stress that the regime of validity of the Boltzmann equation is $k\lambda_{\mathrm{F}} \ll  1$.    Therefore, unless $\alpha$  is  parametrically small,  the long-range Coulomb  interactions will dominate the dynamics of the $\pm 1$ harmonics.   This is why we stated that plasmons  destroy sound waves in typical Fermi liquids, in the introduction.

%Finally, we note that if $\omega=0$,  then any solutions (\ref{eq:anfourier}) are \emph{almost equivalent}  to  a solution of the same equations with $\alpha=0$.  The  only discrepancy between  $\alpha=0$ and $\alpha \ne 0$ is a nonlocal transformation  of $a_0$:   \begin{equation}
%\tilde{a}_0(\mathbf{k})  = a_0(\mathbf{k}) \left(1+\frac{2\mpi \alpha}{\lambda_{\mathrm{F}}|\mathbf{k}| (1+\mathcal{F}_0)}\right).
%\end{equation}
%The variables $\lbrace \ldots, a_{-1}, \tilde{a}_0, a_1, \ldots\rbrace  $  obey the equations of motion (\ref{eq:anfourier}) with  $\alpha=0$.   The effects of long-range Coulomb interactions are then \emph{invisible} to any direct current experiment \cite{lucas3, lucasreview17}.

\section{A Solvable Toy Model}
To find the exact solutions advertised in the introduction, we must  now choose a simple model for  $\mathcal{F}_n$ and $\gamma_n$.   A  simple  solvable model is the relaxation time  model of \cite{levitov1607, lucas1612, levitov1612}: \begin{equation}
\gamma_n = \left\lbrace \begin{array}{ll} 0 &\ n=0 \\ \gamma_{\mathrm{imp}} &\ |n|=1 \\ \gamma &\ |n| \ge 2 \end{array}\right..   \label{eq:solvegamma}
\end{equation}
While  this model is not microscopically realistic when  only two-body collisions are  important \cite{ledwith1, ledwith2}, it captures many non-trivial features of the hydrodynamic to ballistic crossover, and for our purposes   this will suffice.
We will also consider the simplest possible non-trivial choice of $\mathcal{F}_n$:  \begin{equation}
\mathcal{F}_n = \left\lbrace \begin{array}{ll} 0 &\ n \ne 0 \\ \mathcal{F}_0 &\ n=0 \end{array}\right..  \label{eq:solveF}
\end{equation}
These assumptions can be relaxed at the expense of losing some of the analytic tractability.  For a more microscopic model of $\mathcal{F}_n$ and $\gamma_n$,  it would be straightforward to numerically compute the normal modes of (\ref{eq:anfourier}).

%As we have already shown, it suffices to solve this model when $\alpha = 0$.   Using (\ref{eq:F0plasmon}), we can immediately find  results at finite $\alpha$.  

The solvability of (\ref{eq:anfourier}), in the model (\ref{eq:solvegamma}) and (\ref{eq:solveF}), comes from the  following observation.   For $|n|\ge 2$, we have the generic equations \begin{equation}
(\gamma - \mathrm{i}\omega)a_n + \frac{\mathrm{i}kv}{2} (a_{n-1}+a_{n+1})  = 0.  \label{eq:largen}
\end{equation}
Let us look for solutions of the form $a_n = \lambda^{n-1}a_1$ for $n>1$.   (\ref{eq:largen}) implies \begin{equation}
\lambda^2 + \frac{2(\gamma-\mathrm{i}\omega)}{\mathrm{i}kv} \lambda  +1  = 0.
\end{equation}
This is solved by \begin{equation}
\lambda =  -\mathrm{i}\frac{\gamma-\mathrm{i}\omega}{kv} \left( \pm \sqrt{1+\left(\frac{kv}{\gamma-\mathrm{i}\omega}\right)^2}- 1\right) .\label{eq:lambda}
\end{equation}
It is important to keep in mind that $|\lambda| \le 1$ is required in order for a normal mode to exist (be normalizable).

Now, we look for the spectrum of normal modes.   The equations (\ref{eq:anfourier}) are analogous to the solution of tight-binding models in one dimension, with scattering off of a defect near  the  origin.   Looking for solutions with $\lambda = \mathrm{e}^{-\mathrm{i}\phi}$, which heuristically corresponds to a $\mdelta$ function fluctuation of the Fermi surface:  $\Phi(\theta) \sim \mdelta(\theta -  \phi)$, we expect a continuum of normal modes with 
\begin{equation}
\omega = -\mathrm{i}\gamma +  kv\cos\phi  \label{eq:contomega}
\end{equation}
for (almost) any real $0< \phi  \le 2\mpi $.    It remains  to satisfy the boundary conditions  $a_{\pm 1}=0$ ($a_0=0$ then trivially follows).    This  can be done as follows.   With the exception of $\phi = 0,\mpi$,  there are two $\phi$ which have the same value of $\omega$, but different $\lambda$.   We add these two modes together, with suitably chosen constant prefactors so that $a_{\pm 1}$ vanishes.  The final result is
\begin{equation}
a_n = \left\lbrace \begin{array}{ll}   \mathrm{e}^{\mathrm{i}(n-1)\phi} - \mathrm{e}^{\mathrm{i}(1-n)\phi} &\  n>1 \\ 0 &\ n\le 1 \end{array}\right.,
\end{equation}
and a similar mode with  $a_n \ne 0$ only for $n<-1$.   

%We note that there are three ``missing modes" from the above  spectrum.    At $\phi = 0, \mpi$, we cannot find a normal mode with dispersion relation (\ref{eq:contomega}).  Similarly, the modes $\phi =  \mpi/2$ and $\phi = 3\mpi/2$ cannot be distinguished.   We will see that these three modes exactly correspond to the ``hydrodynamic"  modes  (with a finite $\gamma_{\mathrm{imp}}$, only one is truly hydrodynamic).

Now let us  look for the remaining modes, which do not obey (\ref{eq:contomega}).  These are analogous to the ``bound states" of the tight-binding model and will have $a_0$ and/or $a_{\pm 1}$ non-vanishing.  The equations for $a_0$ and $a_{\pm1}$ are \begin{subequations}\begin{align}
-\mathrm{i}\omega a_0 + \frac{\mathrm{i}kv}{2}(a_1+a_{-1}) &= 0,  \label{eq:a0eq} \\
(\gamma_{\mathrm{imp}}-\mathrm{i}\omega )a_{\pm 1} + \frac{\mathrm{i}kv}{2}(a_0+a_{\pm 2}) + \frac{\mathrm{i}kv\mathcal{F}_0}{2} a_0 &= 0.
\end{align}\end{subequations}
  First, we look  for a  mode which has $a_0=0$, but $a_{\pm 1} \ne 0$.    (\ref{eq:a0eq}) implies  that $a_1=-a_{-1}$.   So let us solve just for the modes  with $n>0$.  Making the ansatz $a_n = \lambda^{n-1}a_1$, we find \begin{equation}
  \mathrm{i}\left(\omega + \mathrm{i}\gamma_{\mathrm{imp}}- \frac{ \lambda}{2} kv\right) a_1 = 0.  \label{eq:simplediff}
  \end{equation}
  We find that (\ref{eq:lambda}) and (\ref{eq:simplediff}) are exactly solved by \begin{equation}
\omega  = -\mathrm{i}\gamma_{\mathrm{imp}}  -\mathrm{i}\frac{ v^2}{4(\gamma-\gamma_{\mathrm{imp}})}k^2.  \label{eq:shearmode}
  \end{equation}
  This is the hydrodynamic shear diffusion mode, which has become gapped by momentum relaxation.   In order for the diffusion constant to be positive, and for the theory to be stable, we require that $\gamma>\gamma_{\mathrm{imp}}$.   Also, note that this mode disappears into the continuum  of normal modes once $|\lambda| = 1$.  We  find that this occurs when $\omega = -\mathrm{i}\gamma$, or when $|k|v = 2(\gamma-\gamma_{\mathrm{imp}})$.   Finally, observe that if $\gamma_{\mathrm{imp}}=0$, (\ref{eq:shearmode}) is nothing more than  the conventional hydrodynamic diffusive mode associated with transverse momentum.   The diffusion constant and the attenuation constant of first sound (\ref{eq:puresound}) are  related by viscous hydrodynamics \cite{levitov1607, lucas1612}. 
  
  The final mode has $a_0 \ne 0$, and $a_1 = a_{-1}$.   We find the equations \begin{subequations}\begin{align}
  \omega a_0 &= kva_1, \\
  \left(\omega +\mathrm{i}\gamma_{\mathrm{imp}} - \frac{\lambda}{2}kv\right)a_1 &=  \frac{k(v+\mathcal{F}_0)}{2} a_0,
  \end{align}\end{subequations}
  or \begin{equation}
  \omega  + \mathrm{i}\gamma_{\mathrm{imp}} - \frac{\lambda}{2}kv = \frac{k^2v^2(1+\mathcal{F}_0)}{2\omega}.   \label{eq:simplesound}
  \end{equation}
 (\ref{eq:lambda}) and (\ref{eq:simplesound}) do not have a simple analytic solution for all $\omega$.   Let us instead focus on the two asymptotic  limits of interest for large and small $k$.   We start with the hydrodynamic limit of small $k$,  where we find from (\ref{eq:lambda}) that \begin{equation}
  \lambda \approx -\frac{\mathrm{i}kv}{2(\gamma-\mathrm{i}\omega(k=0))}.
  \end{equation}
  In the special case $\gamma_{\mathrm{imp}}=0$, then $\omega(k=0)=0$ and we find (\ref{eq:puresound}).
  This is the dispersion relation for a first sound wave.    Using the fact that the dynamical viscosity of this toy model is \cite{levitov1607, lucas1612, levitov1612}
  \begin{equation}
  \nu = \frac{v^2}{4\gamma},  \label{eq:viscFL}
  \end{equation}
  we see that the decay rate of sound is consistent with the hydrodynamic coefficients of this model.   
  
Upon setting $\gamma \sim T^2$, one then finds from (\ref{eq:viscFL}) that the viscosity of the Fermi liquid is proportional to $T^{-2}$.   In fact, the viscosity of a Fermi liquid is modified by factors of $\log (T/T_{\mathrm{F}})$ \cite{novikov}.   These additional logarithmic factors are associated with the breakdown of the relaxation time approximation in a two-dimensional Fermi liquid.  More generally, $\nu = v^2/4\gamma_2$, with $\gamma_2$ defined in (\ref{eq:Cn}), and the logarithms observed in the viscosity are due to $\gamma_2$.   Further discussion of the breakdown of the relaxation time approximation  can be found in \cite{ledwith1, ledwith2}.
  \color{black}
  
%  , with the speed of sound \begin{equation}
%  v_{\mathrm{s1}} = \sqrt{\frac{1+\mathcal{F}_0}{2}} v,
%  \end{equation}
%  and the decay rate proportional to half of the dynamical viscosity \begin{equation}
%  .
%  \end{equation}
  
  In the case $\gamma_{\mathrm{imp}} \ne 0$, a bit more care is required.   We expect that in  the $k\rightarrow 0$ limit, there is one gapped mode with $\omega = -\mathrm{i}\gamma_{\mathrm{imp}} + \cdots $, and one diffusive  gapless mode.    If $\gamma_{\mathrm{imp}} \ll \gamma$, then $\gamma - \mathrm{i}\omega(k=0) \approx \gamma$  in both cases and (\ref{eq:puresound}) generalizes  to 
 \begin{equation}
  \omega \approx \pm \sqrt{\frac{v(v+\mathcal{F}_0)}{2} k^2 - \left(\frac{\gamma_{\mathrm{imp}}}{2}+\frac{v^2}{8\gamma}k^2\right)^2  }  - \mathrm{i}\left(\frac{\gamma_{\mathrm{imp}}}{2}+\frac{v^2}{8\gamma}k^2\right)+ \cdots  \label{eq:firstdecay}
  \end{equation}
  If $\gamma_{\mathrm{imp}}$  is comparable  to $\gamma$, then (\ref{eq:firstdecay})  cannot be trusted.    There will be no appreciable first sound mode, and instead we find the following low $k$ expansion of a diffusive mode and a gapped mode:   \begin{subequations}\begin{align}
  \omega  &= -\mathrm{i} \frac{v^2(1+\mathcal{F}_0)}{2\gamma_{\mathrm{imp}}}  k^2 + \mathrm{O}(k^4), \\
  \omega &=  -\mathrm{i}\gamma_{\mathrm{imp}} -  \mathrm{i} \frac{v^2}{2} \left[\frac{1}{\gamma-\gamma_{\mathrm{imp}}}-\frac{1+\mathcal{F}_0}{\gamma_{\mathrm{imp}}} \right]k^2 + \mathrm{O}(k^4).
  \end{align}\end{subequations}
  
At large $k$, we find zero sound waves.  Making the ansatz that \begin{equation}
\omega \approx v_{\mathrm{s0}} k + \zeta + \mathrm{O}\left(k^{-1}\right),
  \end{equation} we find that \begin{equation}
  \lambda \approx \frac{v_{\mathrm{s0}} -\sqrt{v_{\mathrm{s0}} ^2-v^2}}{v} \left[1-\frac{\mathrm{i}\gamma+\alpha}{\sqrt{v_{\mathrm{s0}}^2-v^2}k}+\mathrm{O}\left(\frac{1}{k^2}\right)\right].
  \end{equation}
  To solve for the speed of zero sound waves, $v_{\mathrm{s0}}$, we can neglect $\zeta$.   (\ref{eq:simplesound}) gives \begin{equation}
  v_{\mathrm{s0}} \left(v_{\mathrm{s0}}  + \sqrt{v_{\mathrm{s0}}^2-v^2}\right) = v^2(1+\mathcal{F}_0),
  \end{equation}
  which is solved by \begin{equation}
  v_{\mathrm{s0}}  = \pm v\frac{1+\mathcal{F}_0}{\sqrt{1+2\mathcal{F}_0}}.   \label{eq:vs0}
  \end{equation}
  Next, we can perturbatively solve (\ref{eq:simplesound}) for $\zeta$, and we find \begin{equation}
  \zeta = -\mathrm{i} \frac{(\gamma  + 2\mathcal{F}_0\gamma_{\mathrm{imp}})(1+\mathcal{F}_0)}{(1+2\mathcal{F}_0)^2}.  \label{eq:zerodecay}
  \end{equation}
  The decay rate of zero sound modes is thus approximately set by $\gamma$, but is smaller whenever $\mathcal{F}_0 \ne 0$.   Combining (\ref{eq:vs0}) and (\ref{eq:zerodecay}) we obtain (\ref{eq:zerosound}), when $\gamma_{\mathrm{imp}}=0$.     
  
  We can qualitatively estimate the crossover between the zero and first sound waves by asking when $|\omega| \sim \gamma$.   This occurs when \begin{equation}
k \sim \frac{\gamma}{v\sqrt{1+\mathcal{F}_0}}.
  \end{equation}
As emphasized previously, we are most interested in theories where (the effective) $\mathcal{F}_0 \gg 1$.  In such systems, we find that the hydrodynamic regime -- as measured by the presence of first sound -- is pushed to much longer length scales than naively anticipated.   
  
  If $\gamma_{\mathrm{imp}} \ne 0$, then we find that the zero sound decay rate is approximately shifted by \begin{equation}
  \mathrm{Im}(\omega;  \gamma_{\mathrm{imp}}\ne 0) \approx \mathrm{Im}(\omega;  \gamma_{\mathrm{imp}}= 0) - \gamma_{\mathrm{imp}}  \frac{2\mathcal{F}_0(1+\mathcal{F}_0)}{(1+2\mathcal{F}_0)^2}.  \label{eq:zerosoundimp}
  \end{equation}
In the  limit $\mathcal{F}_0 \gg 1$,  the zero sound decay rate remains approxiamtely offset by (\ref{eq:soundoffset}), analogous  to first sound waves.   When $\mathcal{F}_0$ is small,  the decay rate of first sound waves is, interestingly, nearly independent of momentum relaxation.   We emphasize that this is a theoretical point of interest, as in a real electronic Fermi liquid, the long-range Coulomb interactions always make the effective $\mathcal{F}_0 \gg 1$.

\section{Conclusion}
We have described an analytically solvable toy model for the crossover between sound waves and plasmons in a strongly interacting two dimensional  Fermi liquid.    At long wavelengths, the conventional plasmon mode arising from the long-range Coulomb interactions overtakes both the first sound and zero sound modes of simple two dimensional Fermi liquids.  We have further computed the decay rates  of plasmons  at both short and long wavelengths, and shown that  signatures of hydrodynamics in  the  plasmonic decay are severely limited.   The transition  out of the hydrodynamic regime occurs (for plasmons) at a  parametrically long length scale (\ref{eq:plasmontrans}).    Given that the plasmonic decay rate is also affected by impurities, this means that the electron-impurity scattering rate needs not simply obey $\gamma_{\mathrm{imp}} \ll \gamma$, but \begin{equation}
\gamma_{\mathrm{imp}} \ll \gamma^3 \left(\frac{\lambda_{\mathrm{F}}}{v\alpha}\right)^2,  \label{eq:gammaimp4}
\end{equation}
in  order for impurity scattering to not dominate the decay of plasmons.   While $\gamma_{\mathrm{imp}}  \ll \gamma$ is achievable in 2DEGs, this stronger constraint may not be.  \color{black}

The absence of a sound-like collective mode in 2D electron liquids is completely consistent with the system being in the hydrodynamic regime since, as we show explicitly, the long-range Coulomb interaction modifies all collective modes (both zero and first sound modes of neutral short-range interacting models) to being plasmon-like, with $\omega \propto \sqrt{k}$ at long wavelengths.  Hydrodynamic effects do manifest themselves in the effective plasmon decay rate and in the higher-order wavenumber corrections to the mode dispersion, but these are not easy effects to measure experimentally.  We have thus resolved the conundrum of why strongly interacting and ultraclean 2D Fermi liquids (e.g. graphene, high-mobility 2D GaAs systems) only manifest plasma type long wavelength collective mode dispersion.  This is an essential effect of the long-range nature of Coulomb interaction -- the sound modes are no longer linear in wavenumber in a Coulomb system.  The same behavior should also apply to 3D Coulomb Fermi liquids where the hydrodynamic zero and first sound modes will in fact develop a mass gap because of the long-range nature of the 3D Coulomb interaction, but we know of no examples of 3D Coulomb Fermi liquids  crossing over to the hydrodynamic regime (3D metals are always in the collisionless regime).  Several 2D Coulomb Fermi liquid systems, however, are expected to be in the hydrodynamic regime at moderate temperatures (e.g. monolayer and bilayer graphene, high-mobility 2D GaAs systems), and we predict that the only observable collective modes in these systems will always manifest the square-root in wavenumber long wavelength 2D plasmon dispersion, no matter how deep they are in the hydrodynamic regime (i.e. no matter how strong the quasiparticle colision rate is).  Our work also indicates that the easiest way to see the hydrodynamic sound mode in a 2D system would be to screen out the long-range part of the inter-particle Coulomb interaction by putting a parallel metal gate close to the 2D layer.  In the hydrodynamic regime this will lead to a linearly dispersing sound wave whose speed of sound is sensitive to the properties of the gate.    Sound waves in 2D can also be observed in systems with explicitly short-ranged interactions, such as thin films of normal He-3.

Many previous works have attempted to include the interplay of both electron-electron interactions \cite{singwi, sarma0102, glazman2004} and impurities \cite{sarma9602} in the plasmon dispersion relation of a weakly interacting metal.   Our work provides a simple model, treating impurity scattering and interactions on an equal  footing all  the  way from the  collisionless to the hydrodynamic regime.   In particular, the viscous plasmonic decay rate given in (\ref{eq:plasmon1}) may be much larger than the plasmon decay rate due to electron-phonon scattering \cite{glazman2004}.   More recent work on plasmons in the hydrodynamic regime includes \cite{basovrmp, mirlin2015, svintsov1710}.   Our key findings, including the novel collisionless dispersion relation (\ref{eq:plasmon2}) and the unexpectedly early crossover from hydrodynamic to collisionless plasmons at the length scale (\ref{eq:plasmontrans}), are not contained in these earlier works.    Our work also explains why the collective modes  of 2D Fermi liquids which are expected to be in a hydrodynamic regime, including doped graphene \cite{Sarma:2007ej} and 2D semiconductor systems,  have exhibited  no hints of hydrodynamics.  

We should also make some comments on the implication of our work in the context of the widely used `hydrodynamic dielectric function' approach to collective modes in metals and semiconductors, particularly in calculating the response of finite systems, including  surfaces, interfaces and inhomogeneous electron systems, to external electromagnetic fields.  Some representative, but by no means exhaustive, references are \cite{fbloch, ajbennett, heinrichs1973, fetter1973, eguiluz, barton1979, sarma1979, schiach, sarma1982, sarma1987, yuluo, ciraci}.  This theory is extensively used in the emergent field of nanoplasmonics, where collective electronic response properties of ultrasmall systems are studied for potential technological applications.  This heuristic approach, dating back to Bloch in 1933 \cite{fbloch}, treats the metal (or doped semiconductor) as a hydrodynamic fluid rather uncritically without any consideration for whether the system is or is not in the collision-dominated regime.  In fact, most 3D metals and doped semiconductors are not in the collision-dominated regime since the electron-electron scattering rate is invariably weaker than electron-impurity and electron-phonon scattering rates at low and high temperatures, respectively.  But the simplicity of the hydrodynamic approach makes it particularly attractive for response calculations as a dynamical generalization of the static Thomas-Fermi approximation,  with the fluid pressure term approximated by a parametrized ansatz so as to give the correct plasma dispersion (e.g. as obtained by the random phase approximation) in the bulk, up to second order in wave number.  Such a hydrodynamic response theory has been used extensively for calculating surface and interface plasma modes of finite electron systems in many situations, mainly because of its manifest simplicity and tractability:  the bulk hydrodynamic dielectric function has a simple finite frequency pole with a well-defined analytic form for the wave number dependence of the pole.  Our work shows that such hydrodynamic response theories, while being easy to implement numerically, are not rigorous from a fundamental microscopic perspective.  In particular, a simple ansatz for the fluid pressure applicable for all frequencies in the response calculation is not justifiable.    The dissipative response of the system changes qualitatively between the collision dominated regime at low frequency (where hydrodynamics is valid) to a collisionless ballistic theory at high frequency where hydrodynamics manifestly fails.  Our work also shows that, although the leading order collective mode is always defined by the standard plasma frequency in Coulomb systems (independent of collision-dominated hydrodynamic or collisionless ballistic regime), all higher-order dispersion corrections depend explicitly on whether the system is or is not in the hydrodynamic regime (analogous to the difference between zero sound and first sound).  Since most of these hydrodynamic theories of electron response are focused on obtaining the collective mode dispersion beyond the long wavelength limit (i.e. the so-called `non-local' effects), we caution against taking the quantitative predictions of these theories too literally.  In addition, our work shows that the nature of the collective mode damping depends crucially on whether the system is the collision-dominated hydrodynamic or collisionless ballistic regime, another subtlety not considered in these hydrodynamic response  theories.

%The momentum diffusion mode (\ref{eq:shearmode})  is detectable whenever $\gamma_{\mathrm{imp}} \ll \gamma$, in contrast to (\ref{eq:gammaimp4}).    This  is consistent with the fact that, to  date,  all experimentally observed  hydrodynamic electron flows in Fermi liquids have been direct current transport measurements, consistent with the incompressible Navier-Stokes equations \cite{bandurin, mackenzie, levitov1703}.    

%Given a microscopic calculation of Landau's Fermi liquid parameters $\mathcal{F}_n$, and the collision integral, for a specific material, numerically finding the eigenmodes of (\ref{})  is straightforward.   This may lead to quantitative predictions for 

\addcontentsline{toc}{section}{Acknowledgements}
\section*{Acknowledgements}
AL was supported by the Gordon and Betty Moore Foundation's EPiQS Initiative through Grant GBMF4302.   SDS was supported  by Laboratory for Physical Sciences.

\begin{appendix}
\end{appendix}

\bibliographystyle{unsrt}
\addcontentsline{toc}{section}{References}
\bibliography{soundbib}

\begin{thebibliography}{10}

\bibitem{bandurin}
D.~A.~Bandurin \emph{et al.}
\newblock ``Negative local resistance due to viscous electron backflow in
  graphene",
  \href{http://science.sciencemag.org/content/351/6277/1055}{\textsl{Science}
  \textbf{351} 1055 (2016)},
  \href{http://arxiv.org/abs/1509.04165}{\texttt{arXiv:1509.04165}}.

\bibitem{crossno}
J.~Crossno \emph{et al.}
\newblock ``Observation of the Dirac fluid and the breakdown of the
  Wiedemann-Franz law in graphene",
  \href{http://science.sciencemag.org/content/351/6277/1058}{\textsl{Science}
  \textbf{351} 1058 (2016)},
  \href{http://arxiv.org/abs/1509.04713}{\texttt{arXiv:1509.04713}}.

\bibitem{mackenzie}
P.~J.~W. Moll, P.~Kushwaha, N.~Nandi, B.~Schmidt, and A.~P. Mackenzie.
\newblock ``Evidence for hydrodynamic electron flow in $\mathrm{PdCoO}_2$",
  \href{http://science.sciencemag.org/content/351/6277/1061}{\textsl{Science}
  \textbf{351} 1061 (2016)},
  \href{http://arxiv.org/abs/1509.05691}{\texttt{arXiv:1509.05691}}.

\bibitem{levitov1703}
R.~Krishna~Kumar \emph{et al}.
\newblock ``Super-ballistic flow of viscous electron fluid through graphene
  constrictions", \href{https://doi.org/10.1038/nphys4240}{\textsl{Nature
  Physics} \textbf{13} 1182 (2017)},
  \href{http://arxiv.org/abs/1703.06672}{\texttt{arXiv:1703.06672}}.

\bibitem{lucasreview17}
A.~Lucas and K.~C. Fong.
\newblock ``Hydrodynamics of electrons in graphene",
  \href{http://arxiv.org/abs/1710.08425}{\texttt{arXiv:1710.08425}}.

\bibitem{andreev}
A.~V. Andreev, S.~A. Kivelson, and B.~Spivak.
\newblock ``Hydrodynamic description of transport in strongly correlated
  electron systems",
  \href{http://journals.aps.org/prl/abstract/10.1103/PhysRevLett.106.256804}{\textsl{Physical
  Review Letters} \textbf{106} \texttt{256804} (2011)},
  \href{http://arxiv.org/abs/1011.3068}{\texttt{arXiv:1011.3068}}.

\bibitem{lucas3}
A.~Lucas, J.~Crossno, K.~C. Fong, P.~Kim, and S.~Sachdev.
\newblock ``Transport in inhomogeneous quantum critical fluids and in the Dirac
  fluid in graphene",
  \href{http://journals.aps.org/prb/abstract/10.1103/PhysRevB.93.075426}{\textsl{Physical
  Review} \textbf{B93} \texttt{075426} (2016)},
  \href{http://arxiv.org/abs/1510.01738}{\texttt{arXiv:1510.01738}}.

\bibitem{polini}
I.~Torre, A.~Tomadin, A.~K. Geim, and M.~Polini.
\newblock ``Non-local transport and the hydrodynamic shear viscosity in
  graphene",
  \href{http://journals.aps.org/prb/abstract/10.1103/PhysRevB.92.165433}{\textsl{Physical
  Review} \textbf{B92} \texttt{165433} (2016)},
  \href{http://arxiv.org/abs/1508.00363}{\texttt{arXiv:1508.00363}}.

\bibitem{levitovhydro}
L.~Levitov and G.~Falkovich.
\newblock ``Electron viscosity, current vortices and negative nonlocal
  resistance in graphene",
  \href{http://www.nature.com/nphys/journal/v12/n7/full/nphys3667.html}{\textsl{Nature
  Physics} \textbf{12} 672 (2016)},
  \href{http://arxiv.org/abs/1508.00836}{\texttt{arXiv:1508.00836}}.

\bibitem{levitov1607}
H.~Guo, E.~Ilseven, G.~Falkovich, and L.~Levitov.
\newblock ``Higher-than-ballistic conduction of viscous electron flows",
  \href{http://www.pnas.org/content/114/12/3068.abstract}{\textsl{Proceedings
  of the National Academy of Sciences} \textbf{114} 3068 (2017)},
  \href{http://arxiv.org/abs/1607.07269}{\texttt{arXiv:1607.07269}}.

\bibitem{lucas1612}
A.~Lucas.
\newblock ``Stokes paradox in electronic Fermi liquids",
  \href{http://link.aps.org/doi/10.1103/PhysRevB.95.115425}{\textsl{Physical
  Review} \textbf{B95} \texttt{115425} (2017)},
  \href{http://arxiv.org/abs/1612.00856}{\texttt{arXiv:1612.00856}}.

\bibitem{levitov1612}
H.~Guo, E.~Ilseven, G.~Falkovich, and L.~Levitov.
\newblock ``Stokes paradox, back reflections and interaction-enhanced
  conductance",
  \href{http://arxiv.org/abs/1612.09239}{\texttt{arXiv:1612.09239}}.

\bibitem{hartnoll1705}
A.~Lucas and S.~A. Hartnoll.
\newblock ``Kinetic theory of transport for inhomogeneous electron fluids",
  \href{http://arxiv.org/abs/1706.04621}{\texttt{arXiv:1706.04621}}.

\bibitem{lucasRFB}
A.~Lucas.
\newblock ``Kinetic theory of electronic transport in random magnetic fields",
  \href{http://arxiv.org/abs/1710.11141}{\texttt{arXiv:1710.11141}}.

\bibitem{pines}
D.~Pines and P.~Nozi\`eres.
\newblock \emph{The Theory of Quantum Liquids, Volume I}
  \href{http://www.amazon.com/Theory-Of-Quantum-Liquids-Advanced/dp/0201407744}{(W.
  A. Benjamin, 1966)}.

\bibitem{landauvol9}
E.M. Lifshitz and L.~P. Pitaevskii.
\newblock \emph{Statistical Physics Part 2}
  \href{https://www.amazon.com/Statistical-Physics-Part-Theory-Condensed/dp/0080230725/ref=sr_1_1?ie=UTF8&qid=1518635166&sr=8-1&keywords=statistical+physics+part+2}{(Butterworth
  Heinemann, 1980)}.

\bibitem{giulianivignale}
G.~F. Giuliani and G.~Vignale.
\newblock \emph{Quantum Theory of the Electron Liquid}
  \href{https://www.amazon.com/Quantum-Theory-Electron-Gabriele-Giuliani/dp/0521527961/ref=sr_1_1?s=books&ie=UTF8&qid=1518635178&sr=1-1&keywords=giuliani+vignale}{(Cambridge
  University Press, 2008)}.

\bibitem{DS}
M.~Dyakonov and M.~Shur.
\newblock ``Shallow water analogy for a ballistic field effect transistor: New
  mechanism of plasma wave generation by dc current",
  \href{http://journals.aps.org/prl/abstract/10.1103/PhysRevLett.71.2465}{\textsl{Physical
  Review Letters} \textbf{71} 2465 (1993)}.

\bibitem{tauk}
R.~Tauk \emph{et al.}
\newblock ``Plasma wave detection of terahertz radiation by silicon field
  effects transistors: responsivity and noise equivalent power",
  \href{http://scitation.aip.org/content/aip/journal/apl/89/25/10.1063/1.2410215}{\textsl{Applied
  Physics Letters} \textbf{89} \texttt{253511} (2006)}.

\bibitem{giliberti}
V.~Giliberti, A.~Di Gaspare, E.~Giovine, M.~Ortolani, L.~Sorba, G.~Biasiol,
  V.~V. Popov, D.~V. Fateev, and F.~Evangelisti.
\newblock ``Downconversion of terahertz radiation due to intrinsic hydrodynamic
  nonlinearity of a two-dimensional electron plasma",
  \href{http://dx.doi.org/10.1103/PhysRevB.91.165313}{\textsl{Physical Review}
  \textbf{B91} \texttt{165313} (2015)}.

\bibitem{abel}
W.~R. Abel, A.~C. Anderson, and J.~C. Wheatley.
\newblock ``Propagation of zero sound in liquid He$^3$ at low temperatures",
  \href{http://journals.aps.org/prl/abstract/10.1103/PhysRevLett.17.74}{\textsl{Physical
  Review Letters} \textbf{17} 74 (1966)}.

\bibitem{sarma96}
L.~Zheng and S.~Das Sarma.
\newblock ``Coulomb scattering lifetime of a two-dimensional electron gas",
  \href{https://doi.org/10.1103/PhysRevB.89.235431}{\textsl{Physical Review}
  \textbf{B53} 9964 (1996)},
  \href{http://arxiv.org/abs/cond-mat/9602066}{\texttt{arXiv:cond-mat/9602066}}.

\bibitem{novikov}
D.~S. Novikov.
\newblock ``Viscosity of a two-dimensional Fermi liquid",
  \href{http://arxiv.org/abs/cond-mat/0603184}{\texttt{arXiv:cond-mat/0603184}}.

\bibitem{sarma13}
Q.~Li and S.~Das Sarma.
\newblock ``Finite temperature inelastic mean free path and quasiparticle
  lifetime in graphene",
  \href{https://doi.org/10.1103/PhysRevB.87.085406}{\textsl{Physical Review}
  \textbf{B87} \texttt{085406} (2013)},
  \href{http://arxiv.org/abs/1211.6430}{\texttt{arXiv:1211.6430}}.

\bibitem{polini1506}
A.~Principi, G.~Vignale, M.~Carrega, and M.~Polini.
\newblock ``Bulk and shear viscosities of the 2D electron liquid in a doped
  graphene sheet",
  \href{http://journals.aps.org/prb/abstract/10.1103/PhysRevB.93.125410}{\textsl{Physical
  Review} \textbf{B93} \texttt{125410} (2016)},
  \href{http://arxiv.org/abs/1506.06030}{\texttt{arXiv:1506.06030}}.

\bibitem{zala2001}
G.~Zala, B.~N. Narozhny, and I.~L. Aleiner.
\newblock ``Interaction corrections at intermediate temperatures: Longitudinal
  conductivity and kinetic equation",
  \href{https://doi.org/10.1103/PhysRevB.64.214204}{\textsl{Physical Review}
  \textbf{B64} \texttt{214204} (2001)},
  \href{http://arxiv.org/abs/cond-mat/0105406}{\texttt{arXiv:cond-mat/0105406}}.

\bibitem{stern}
T.~Ando, A.~B. Fowler, and F.~Stern.
\newblock ``Electronic properties of two-dimensional systems",
  \href{https://doi.org/10.1103/RevModPhys.54.437}{\textsl{Reviews of Modern
  Physics} \textbf{54} 437 (1982)}.

\bibitem{hwangplasmon}
E.~H. Hwang and S.~Das Sarma.
\newblock ``Dielectric function, screening and plasmons in two-dimensional
  graphene",
  \href{http://journals.aps.org/prb/abstract/10.1103/PhysRevB.75.205418}{\textsl{Physical
  Review} \textbf{B75} \texttt{205418} (2007)},
  \href{http://arxiv.org/abs/cond-mat/0610561}{\texttt{arXiv:cond-mat/0610561}}.

\bibitem{sarmaplasmon}
S.~Das Sarma and E.~H. Hwang.
\newblock ``Collective modes of the massless Dirac plasma",
  \href{https://doi.org/10.1103/PhysRevLett.102.206412}{\textsl{Physical Review
  Letters} \textbf{102} \texttt{206412} (2009)},
  \href{http://arxiv.org/abs/0902.3822}{\texttt{arXiv:0902.3822}}.

\bibitem{levitovsound}
T.~V. Phan, J.~C.~W. Song, and L.~S. Levitov.
\newblock ``Ballistic heat transfer and energy waves in an electron system",
  \href{http://arxiv.org/abs/1306.4972}{\texttt{arXiv:1306.4972}}.

\bibitem{fogler3}
Z.~Sun, D.~N. Basov, and M.~M. Fogler.
\newblock ``Adiabatic amplification of plasmons and demons in 2d systems",
  \href{https://doi.org/10.1103/PhysRevLett.117.076805}{\textsl{Physical Review
  Letters} \textbf{117} \texttt{076805} (2016)},
  \href{http://arxiv.org/abs/1601.02722}{\texttt{arXiv:1601.02722}}.

\bibitem{lucasplasma}
A.~Lucas.
\newblock ``Sound waves and resonances in electron-hole plasma",
  \href{http://journals.aps.org/prb/abstract/10.1103/PhysRevB.93.245153}{\textsl{Physical
  Review} \textbf{B93} \texttt{245153} (2016)},
  \href{http://arxiv.org/abs/1604.03955}{\texttt{arXiv:1604.03955}}.

\bibitem{ledwith1}
P.~Ledwith, H.~Guo, and L.~Levitov.
\newblock ``Fermion collisions in two dimensions",
  \href{http://arxiv.org/abs/1708.01915}{\texttt{arXiv:1708.01915}}.

\bibitem{ledwith2}
P.~Ledwith, H.~Guo, A.~V. Shytov, and L.~Levitov.
\newblock ``Head-on collisions and scale-dependent viscosity in two-dimensional
  electron systems",
  \href{http://arxiv.org/abs/1708.02376}{\texttt{arXiv:1708.02376}}.

\bibitem{singwi}
A.~Czachor, A.~Holas, S.~R. Sharma, and K.~S. Singwi.
\newblock ``Dynamical correlations in a two-dimensional electron gas:
  First-order perturbation theory",
  \href{https://doi.org/10.1103/PhysRevB.25.2144}{\textsl{Physical Review}
  \textbf{B25} 2144 (1982)}.

\bibitem{sarma0102}
E.~H. Hwang and S.~Das Sarma.
\newblock ``Plasmon dispersion in dilute 2D electron systems: Quantum-Classical
  and Wigner Crystal-Electron Liquid Crossover",
  \href{https://doi.org/10.1103/PhysRevB.64.165409}{\textsl{Physical Review}
  \textbf{B64} \texttt{165409} (2001)},
  \href{http://arxiv.org/abs/cond-mat/0102057}{\texttt{arXiv:cond-mat/0102057}}.

\bibitem{glazman2004}
E.~G. Mishchenko, M.~Yu. Reizer, and L.~I. Glazman.
\newblock ``Plasmon attenuation and optical conductivity of a two-dimensional
  electron gas",
  \href{https://doi.org/10.1103/PhysRevB.69.195302}{\textsl{Physical Review}
  \textbf{B69} \texttt{195302} (2004)},
  \href{http://arxiv.org/abs/cond-mat/0312684}{\texttt{arXiv:cond-mat/0312684}}.

\bibitem{sarma9602}
S.~Das Sarma and E.~H. Hwang.
\newblock ``Dynamical response of a one dimensional quantum wire electron
  system", \href{https://doi.org/10.1103/PhysRevB.54.1936}{\textsl{Physical
  Review} \textbf{B54} 1936 (1996)},
  \href{http://arxiv.org/abs/cond-mat/9602157}{\texttt{arXiv:cond-mat/9602157}}.

\bibitem{basovrmp}
D.~N. Basov, M.~M. Fogler, A.~Lanzara, F.~Wang, and Y.~Zhang.
\newblock ``\emph{Colloquium:} Graphene spectroscopy",
  \href{https://doi.org/10.1103/RevModPhys.86.959}{\textsl{Reviews of Modern
  Physics} \textbf{86} 959 (2014)},
  \href{http://arxiv.org/abs/1407.6721}{\texttt{arXiv:1407.6721}}.

\bibitem{mirlin2015}
U.~Briskot, M.~Sch\"utt, I.~V. Gornyi, M.~Titov, B.~N. Narozhny, and A.~D.
  Mirlin.
\newblock ``Collision-dominated nonlinear hydrodynamics in graphene",
  \href{https://doi.org/10.1103/PhysRevB.92.115426}{\textsl{Physical Review}
  \textbf{B92} \texttt{115426} (2015)},
  \href{http://arxiv.org/abs/1507.08946}{\texttt{arXiv:1507.08946}}.

\bibitem{svintsov1710}
D.~Svintsov.
\newblock ``Hydrodynamic-to-ballistic crossover in Dirac fluid",
  \href{http://arxiv.org/abs/1710.05054}{\texttt{arXiv:1710.05054}}.

\bibitem{Sarma:2007ej}
S.~Das Sarma, E.~H. Hwang, and W-K. Tse.
\newblock ``Many-body interaction effects in doped and undoped graphene: Fermi
  liquid versus non-Fermi liquid",
  \href{https://doi.org/10.1103/PhysRevB.75.121406}{\textsl{Physical Review}
  \textbf{B75} \texttt{121406} (2007)},
  \href{http://arxiv.org/abs/cond-mat/0610581}{\texttt{arXiv:cond-mat/0610581}}.

\bibitem{fbloch}
F.~Bloch.
\newblock ``Bremsverm\"ogen von atomen mit mehreren elektronen",
  \href{https://doi.org/10.1007/BF01344553}{\textsl{Zeitschrift f\"ur Physik}
  \textbf{81} 363 (1933)}.

\bibitem{ajbennett}
A.~J. Bennett.
\newblock ``Influence of the electron charge distribution on surface-plasmon
  dispersion", \href{https://doi.org/10.1103/PhysRevB.1.203}{\textsl{Physical
  Review} \textbf{B1} 203 (1970)}.

\bibitem{heinrichs1973}
J.~Heinrichs.
\newblock ``Hydrodynamic theory of surface-plasmon dispersion",
  \href{https://doi.org/10.1103/PhysRevB.7.3487}{\textsl{Physical Review}
  \textbf{B7} 3487 (1973)}.

\bibitem{fetter1973}
A.~L. Fetter.
\newblock ``Electrodynamics of a layered electron gas. I. Single layer",
  \href{https://doi.org/10.1016/0003-4916(73)90161-9}{\textsl{Annals of
  Physics} \textbf{81} 367 (1973)}.

\bibitem{eguiluz}
A.~Eguiluz, S.~C. Ying, and J.~J. Quinn.
\newblock ``Influence of the electron density profile on surface plasmons in a
  hydrodynamic model",
  \href{https://doi.org/10.1103/PhysRevB.11.2118}{\textsl{Physical Review}
  \textbf{B11} 2118 (1975)}.

\bibitem{barton1979}
G.~Barton.
\newblock ``Some surface effects in the hydrodynamic model of metals",
  \href{https://doi.org/10.1088/0034-4885/42/6/001}{\textsl{Reports on Progress
  in Physics} \textbf{42} 963 (1979)}.

\bibitem{sarma1979}
S.~Das Sarma and J.~J. Quinn.
\newblock ``Hydrodynamic model of linear response for a jellium surface:
  Nonretarded limit",
  \href{https://doi.org/10.1103/PhysRevB.20.4872}{\textsl{Physical Review}
  \textbf{B20} 4872 (1979)}.

\bibitem{schiach}
W.~L. Schiach and C.~Schwartz.
\newblock ``Phonons at metal surfaces",
  \href{https://doi.org/10.1103/PhysRevB.25.7365}{\textsl{Physical Review}
  \textbf{B25} 7365 (1982)}.

\bibitem{sarma1982}
S.~Das Sarma.
\newblock ``Electrodynamic response of a bounded electron gas in hydrodynamic
  formalism: Theory and applications",
  \href{https://doi.org/10.1103/PhysRevB.26.6559}{\textsl{Physical Review}
  \textbf{B26} 6559 (1982)}.

\bibitem{sarma1987}
S.~Das Sarma.
\newblock ``Nonlocal theory for surface-plasmon excitation in simple metals",
  \href{https://doi.org/10.1103/PhysRevB.36.3026}{\textsl{Physical Review}
  \textbf{B36} 3026 (1987)}.

\bibitem{yuluo}
Y.~Luo, A.~I. Fernandez-Dominguez, A.~Wiener, S.~A. Maier, and J.~B. Pendry.
\newblock ``Surface plasmons and nonlocality: a simple model",
  \href{https://doi.org/10.1103/PhysRevLett.111.093901}{\textsl{Physical Review
  Letters} \textbf{111} \texttt{093901} (2013)},
  \href{http://arxiv.org/abs/1308.1708}{\texttt{arXiv:1308.1708}}.

\bibitem{ciraci}
C.~Ciraci and F.~Della Sala.
\newblock ``Quantum hydrodynamic theory for plasmonics: impact of the electron
  density tail",
  \href{https://doi.org/10.1103/PhysRevB.93.205405}{\textsl{Physical Review}
  \textbf{B93} \texttt{205405} (2016)},
  \href{http://arxiv.org/abs/1601.01584}{\texttt{arXiv:1601.01584}}.

\end{thebibliography}

\end{document}